\documentclass[nofootinbib,twocolumn,prd,a4paper,final,superscriptaddress,longbibliography]{revtex4-1}

\usepackage{graphicx}
\usepackage{bm}
\usepackage{amsmath,amssymb}

\newcommand{\dd}{{\mathrm{d}}} 
\newcommand{\pd}{{\partial}} 
\newcommand{\DOne}{{\Delta_1}} 
\newcommand{\DTwo}{{\Delta_2}} 
\newcommand{\rmA}{\mathrm{A}} 
\newcommand{\rmB}{\mathrm{B}} 
\newcommand{\rmC}{\mathrm{C}} 

\newcommand{\Kahler}{K$\ddot{\mathrm{a}}$hler }
\newcommand{\Backlund}{B$\ddot{\mathrm{a}}$cklund }

\begin{document}
	
	\title{BPS Sphalerons in the $F_2$ Non-Linear Sigma Model}
	
	\author{Yuki Amari}
	\email{amari.yuki.ph@gmail.com}
	\address{Department of Physics, Tokyo University of Science, Noda, Chiba 278-8510, Japan}
	\address{Instituto de Fisica de S\~ao Carlos; IFSC/USP, Universidade de S\~ao Paulo - USP, Caixa Postal 369, CEP 13560-970, S\~ao Carlos-SP, Brazil}

	\author{Nobuyuki Sawado}
	\email{sawado@ph.noda.tus.ac.jp}
	\address{Department of Physics, Tokyo University of Science, Noda, Chiba 278-8510, Japan}

	\vspace{.5 in}
	\small

	\date{\today}
	
	\begin{abstract}
			
	We construct static and also time-dependent solutions in a non-linear sigma model with target space being 
	the flag manifold $F_2=SU(3)/U(1)^2$ on the four dimensional Minkowski space-time by analytically 
	solving the second order Euler-Lagrange equation. 
	We show the static solutions saturate an energy lower bound and can be derived from coupled first 
	order equations though they are saddle point solutions. 
	We also discuss basic properties of the time-dependent solutions. 
	\end{abstract}
	
	
	\maketitle 
	\section{Introduction}

	Non-linear sigma (NL$\sigma$) models have been extensively studied in various branch of physics. 
	Specifically, the $O(3)$ NL$\sigma$ model in two dimensions has been identified as a good toy model of the $SU(2)$ pure Yang-Mills theory in four dimensions, 
	because they share many important features such as conformal invariance, dynamical mass gap and existence of the instanton solution.
	Moreover, the $O(3)$ NL$\sigma$ model (with higher differential terms) can be derived as a low energy effective theory for 
	the $SU(2)$ Yang-Mills theory \cite{Faddeev:1998eq,Gies:2001hk}, Heisenberg model \cite{Haldane:1983ru} etc..  
	The topological soliton solutions like the instantons and the Hopfions play a crucial role 
	for the non-perturbative aspects in the theories \cite{Polyakov:1975yp,Faddeev:1996zj}. 
	
	It is of great importance to consider $SU(N)$ generalizations of the $O(3)$ NL$\sigma$ model,
	because they share some basic properties with the $SU(N)$ Yang-Mills theory 
	and may possibly be derived from fundamental theories like the $SU(N)$ Yang-Mills theory or the $SU(N)$ Heisenberg model. 
	There may be several variants, and in particular two feasible cases have widely been examined; the $CP^{N-1}$ and  $F_{N-1}$ NL$\sigma$ model.
	The models are NL$\sigma$ models whose target spaces are the complex projective space $CP^{N-1}=SU(N)/U(N-1)$ and the flag space $F_{N-1}=SU(N)/U(1)^{N-1}$, respectively\footnote{For $N\geq4$ there are also intermediate options whose target space is the partial flag manifolds
		$SU(N)/U(1)^m\times S(U(n_1)\times\cdots\times U(n_k))$ where $m+\sum n_i=N-1, k\geq1$ which is including the 
		Grassmannian manifolds.}.
	Note that the $CP^1 (=F_1)$ NL$\sigma$ model is equivalent to the $O(3)$ NL$\sigma$ model at least in classical level.
	Indeed, the $CP^{N-1}$ models arose as an effective theory from the $SU(N)$ Yang-Mills theory with the minimal case of 
	Maximal Abelian gauge (MAG)~\cite{Kondo:2008xa}, ferromagnetic $SU(N)$ Heisenberg model \cite{Affleck:1984ar,Read:1989zz} and so on 
	\cite{Eto:2009bh,Eto:2013hoa,Hindmarsh:1991jq,Garaud:2012pn,Bykov:2010tv}.
	On the other hand, the $F_{N-1}$ model was considered from the $SU(N)$ Yang-Mills theory 
	by means of the maximal case of MAG~\cite{Faddeev:1998yz,Kondo:2008xa}.
	Moreover, the $F_{2}$ model was derived from the $SU(3)$ anti-ferromagnetic Heisenberg model in 1+1 dimensions \cite{Bykov:2011ai,Lajko:2017wif} 
	and in 2+1 dimensions \cite{Bernatska:2009,Shannon:2013}.

	In this paper, we study classical solutions of the $F_{2}$ model.
	There are quite few studies for classical solutions of the $F_{N-1}$ models, even in the $N=3$ case, 
	though ones of the $CP^{N-1}$ models have been studied for all $N$ since the late 70s \cite{DAdda:1978vbw,Din:1980jg}.
	To our knowledge two seminal papers about classical solutions of the $F_{2}$ model are the ones.
	In \cite{Ueda:2016}, it was shown that the BPS equation is solved by the $F_{1}$ instantons embedded into the target space $F_{2}$.
	In \cite{Bykov:2015pka}, the author studied classical solutions of the $F_{2}$ model with an additional term called Kalb-Ramond 
	field which is non topological term and then contributes to the Euler-Lagrange equation.
	In some special choices of the coupling constant, the general static solutions were discussed in context of the integrability \cite{Bykov:2014efa}.
	Note that the Kalb-Ramond field can naturally appear when one derives the $F_{2}$ model from the $SU(3)$ spin chain \cite{Bykov:2011ai,Lajko:2017wif}. 
	In contrast, from the $SU(3)$ Yang-Mills theory or the $SU(3)$ Heisenberg model on lattice, the term may not be provided.
	 
	Main goal of this paper is to find genuine (which means nonembedding) solutions
	in the $F_2$ NL$\sigma$ model without the Kalb-Ramond term.
	The model in (3+1)-dimensional Minkowski space-time is defined by the Lagrangian density 
	\begin{equation}
		\mathcal{L}=
		-\left(\left\|Z_\rmA^\dagger\partial_\mu Z_\rmB \right\|^2
		+\left\|Z_\rmB^\dagger\partial_\mu Z_\rmC \right\|^2
		+\left\|Z_\rmC^\dagger\partial_\mu Z_\rmA \right\|^2\right).
		\label{F2 nls}
	\end{equation}
	Note that, the double vertical line norm for (composite) vector fields denotes the Minkowski norm with the metric $g_{\mu\nu}=(-,+,+,+)$, thus the square of the norm is
	not positive definite. 
	Here, the fields $Z_a ~(a=\rmA,\rmB,\rmC)$ are three components complex vectors which form a complete basis. Therefore we can construct $SU(3)$ matrix from $U(3)$ matrix $U=(Z_\rmA, Z_\rmB, Z_\rmC)$ and the fields satisfy the orthonormal condition and the completeness condition i.e.	
	\begin{gather}
	Z_a^\dagger Z_b=\delta_{ab}~,   \label{orthonormal}
	\\
	Z_\rmA\otimes Z_\rmA^\dagger
	+Z_\rmB\otimes Z_\rmB^\dagger
	+Z_\rmC\otimes Z_\rmC^\dagger=\mathbf{1}_3 \label{completeness}
	\end{gather}
	where $\mathbf{1}_3$ is the $3\times 3$ identity matrix.
	The Lagrangian is invariant under the left global $U(3)$ transformation such as
	$U\rightarrow gU,~ g\in U(3).$
	In addition, it is also invariant under the local $U(1)^3$ transformation i.e. three $U(1)$ phase rotations
	$Z_a\rightarrow Z_a^\prime=e^{i\Theta_a}Z_a$, $a=\rmA,\rmB,\rmC$ .
	
	The paper is organized as follows. 	
	In Sec.II, we briefly review the previous study of an embedding solution of the model. In Sec.III, the 
	basic information for constructing the solutions including the parametrization, the topological charges are given.  
	Sec.IV presents both the static and the time-dependent solutions of the model. Some properties of the energy, for example an energy bound for the static solutions, are devoted in Sec.V. Summary and discussion are in Sec.VI.

	\section{Embedding solutions}
	As we mentioned, the authors of \cite{Ueda:2016} found a BPS solution which is just an $F_1(=CP^1)$ instanton embedded into 
	the target space $F_2$.
	Let us begin with a brief review of their work with introduction of some notations.
		
	The discussion is based on an analogy with the BPS sector in the $CP^2$ model. For sake the reader's understanding, 
	we introduce the covariant derivative
	\begin{equation}
	D^{(a)}_\mu  =\partial_\mu -\left(Z_a^\dagger \partial_\mu Z_a\right) \qquad  a=\rmA,\rmB,\rmC\,.
	\label{Covariant deivative}
	\end{equation}
	and write the static energy per unit length such as
	\begin{align}
	E=\frac{1}{2}\int \dd^2 x
	\sum_{a=\rmA,\rmB,\rmC}
	\left\|D^{(a)}_i Z_a\right\|^2~.
	\label{Energy1}
	\end{align}
	Each term in \eqref{Energy1} can be viewed as the static energy of the  $CP^2$ model linked by the orthonormal condition \eqref{orthonormal}.
	Thus, by simply applying the BPS bound of the $CP^2$ model to the three terms respectively, an energy lower bound may be derived as
	\begin{equation}
	 E\geq \pi\left(|N_{\rm A}|+|N_{\rm B}|+|N_{\rm C}|\right) \label{Bound 1}
	\end{equation} 
	where $N_a$ is the topological degrees of the map $Z_a:\mathbb{R}^2\sim
	S^2\to CP^2$ defined as
	\begin{equation}
	 N_a=\frac{i}{2\pi}\int \dd^2x ~\epsilon^{ij}\left(D^{(a)}_{~i} Z_a\right)^\dagger D^{(a)}_{~j} Z_a~.
	 \label{Na}
	\end{equation}
	Note that there is a constraint among the topological degrees of the form
	\begin{equation}
		N_\rmA+N_\rmB+N_\rmC=0.
		\label{constraint N}
	\end{equation}
	The equality in \eqref{Bound 1} is satisfied if the following (implicitly) coupled first derivative equations are simultaneously satisfied:
	\begin{equation}
	 D^{(a)}_i Z_a=\pm i\epsilon^{ij} D_j^{(a)} Z_a \qquad \forall a=\rmA,\rmB,\rmC.
	 \label{BPS eq}
	\end{equation}
	Unfortunately, the solutions of \eqref{BPS eq} are given by only a very limited class of configurations, i.e.
	an $F_1$ instantons embedded into the target space $F_2$, for which one of the vectors $Z_a$ is fixed as a trivial one.
	By performing a proper global transformation, the solutions can generally be written as
	\begin{equation}
	\begin{split}
	&
	Z_\rmA=\dfrac{1}{\sqrt{\Delta}}
	\left(
	1,
	p(z)/q(z),
	0
	\right)^T,
	\\
	&
	Z_\rmB=\dfrac{1}{\sqrt{\Delta}}
	\left(
	p(\bar{z})/q(\bar{z}),
	-1,
	0
	\right)^T,
	\\	
	&
	Z_\rmC=
	\left(
	0, 
	0, 
	1
	\right)^T,
	\end{split}
	\label{embedding sol}
	\end{equation}
	with $\Delta\equiv 1+|p(z)|^2/|q(z)|^2$. Here, we have 
	introduced the standard complex variables
	 \begin{equation}
	 z=x^1+ix^2 ,\qquad \bar{z}=x^1-ix^2, 
	 \end{equation}
	 and the functions $p(z)$ and $q(z)$ which are polynomials of $z$ with no common roots.  
	 The solution \eqref{embedding sol} has apparently $N_\rmC=0$,  
	 and therefore, the energy is given by
	 \begin{equation}
	 	E=\pi(|N_\rmA|+|N_\rmB|)=2\pi N_\rmA=-2\pi N_\rmB
	 \end{equation}
	 where $N_A(=-N_B)$ is given the highest power of $z$ in either $p(z)$ or $q(z)$.
	 
	 The embedding solutions are unique possibility of solutions in the equation \eqref{BPS eq}.
	 For, \eqref{BPS eq} requires all the vectors $Z_a$ are holomorphic or anti-holomorphic, 
	 but there is no pair of three components (anti-) holomorphic vectors without constant vectors.
	 If one wants to get more general solution, obviously different strategy has to be implemented. 
	 Our approach for the problem is to solve the Euler-Lagrange equation, without touching first-order equations. 
	 For the solutions, of course, the energy bound \eqref{Bound 1} is no longer saturated.

	 The notable thing is that the nonembedding solutions satisfy a new coupled first-order equations different from \eqref{BPS eq}.
	 As a result, we obtain new energy lower bound which of course is different from \eqref{Bound 1}.
	 Note that the energy is higher than that of embedding solutions 
	 and the new energy bound contains a non-topological term. 
	 The detailed analysis will be thoroughly discussed in Sec.V.   
	 Existence of such solutions will give us deeper understanding of the BPS structure of NL$\sigma$ models whose target space is non-symmetric space. 	
		
	\section{Preliminaries}
	This section is preliminaries for analyzing the Euler-Lagrange equations and some properties of the energy functional. 
	
	\subsection{Parametrization}
      In order to simplify the analysis, we parametrize the vectors $Z_a$ in terms of complex scalar fields \cite{Picken:1988fw}.
      It can be realized by means of the isomorphism
      \begin{equation}
      SU(3)/U(1)^2\cong SL(3,\mathbb{C})/B_{+}
      \end{equation}
      where $B_{+}$ is the Borel subgroup of the upper triangular matrices with determinant being equal to one. 
      In geometrical point of view, the complex scalar fields mean the local coordinates of the manifold $F_2$.
      The coordinates can be introduced though a mapping $SU(3)/U(1)^2 \to SL(3,\mathbb{C})/B_{+}$, which is an extension of the stereographic projection i.e. $S^2 \to \mathbb{R}^2$. 
	By a formal result of the map, one can write an element of $F_2$ in terms of local coordinates $u_i, i=1,2,3$ as the lower triangular matrix
	\begin{equation}
	X=\left(\begin{array}{ccc}
	1 & 0 & 0 \\ 
	u_1 & 1 & 0 \\ 
	u_2 & u_3 & 1
	\end{array} \right)\,.
	\label{low tri}
	\end{equation}
	Since $\det X=1$, $X$ is an element of $SL(3,\mathbb{C})$.
	Note that, however, it is not necessary for $X$ to be an unitary matrix.
	
	Next, in order to obtain the $SU(3)$ matrix $U$ or the vectors $Z_a$ parametrized in terms of the scalar fields, we consider the inverse of the mapping, i.e. $SL(3,\mathbb{C})/B_{+}\to SU(3)/U(1)^2$.
	Then, one can construct $U$ from $X$ by means of the so-called Iwasawa decomposition:
	any element of $SL(3,\mathbb{C})$ may be factorized into an element of $SU(3)$ and $B_{+}$ 
	in a unique fashion, up to torus elements of $U(1)^2$.
	The procedure is as following.
	We regard $X$ as element of $SL(3,\mathbb{C})$ which can be expressed in terms of the column vectors such as 
	\begin{equation}
	X=(c_1,c_2,c_3)\in SL(3,\mathbb{C}) 
	\end{equation}
	with
	\begin{equation}
	c_1=(1,u_1,u_2)^T~, \quad 
	c_2=(0,1,u_3)^T~,\quad 
	c_3=(0,0,1)^T .\label{cvec}
	\end{equation}
	In this case, the Iwasawa decomposition can be proved by the Gramm-Schmit orthogonalization process for \eqref{cvec}. Then one obtains mutually orthogonal vectors 
	\begin{equation}
		\begin{split}
		& e_\rmA=c_1~,
		\\
		& e_\rmB=c_2-\frac{(c_2,e_\rmA)}{(e_\rmA,e_\rmA)}e_\rmA~, 
		\\
		&
		e_\rmC=c_3-\frac{(c_3,e_\rmB)}{(e_\rmB,e_\rmB)}e_\rmB-\frac{(c_3,e_\rmA)}{(e_\rmA,e_\rmA)}e_\rmA
		\end{split}
		\label{vector e}
	\end{equation}
	where the inner product is defined as
	\begin{equation}
	(c_i,c_j)=c_j^\dagger c_i ~.
	\end{equation}
	By normalizing the vectors \eqref{vector e}, one obtains an orthonormal basis. Therefore, we can define $Z_a=\frac{1}{\sqrt{(e_a,e_a)}}e_a$, 
	$a=\rmA, \rmB, \rmC$, and they can explicitly be written as
	\begin{equation}
	\begin{split}
	Z_\rmA&=
	\frac{1}{\sqrt{\Delta_1}}
	\left(\begin{array}{c}
	1\\ 
	u_1\\ 
	u_2
	\end{array} \right),
	\\
	Z_\rmB&=
	\frac{1}{\sqrt{\Delta_1\Delta_2}}
	 \left(\begin{array}{c}
	-u_1^*-u_2^*u_3\\
	1-u_1u_2^*u_3+|u_2|^2\\
	-u_1^*u_2+u_3+u_3|u_1|^2 
	\end{array} \right),
	\\
	Z_\rmC&=
	\frac{1}{\sqrt{\Delta_2}}
	\left(\begin{array}{c}
	u_1^*u_3^*-u_2^*\\
	-u_3^*\\
	1 
	\end{array}
	\right)
	\end{split}
	\label{explict Z}
	\end{equation}
	where 
	\begin{equation}
	\begin{split}
	\Delta_1&=1+|u_1|^2+|u_2|^2~,
	\\
	\Delta_2&=1+|u_3|^2+|u_1u_3-u_2|^2.
	\end{split}
	\label{formdelta}
	\end{equation}
	
	For later convenience, we express the off-diagonal components of the one-form $Z_a^\dagger\dd Z_b$ as
	\begin{equation}
	\begin{split}
	Z_\rmB^\dagger\dd Z_\rmA=&\frac{1}{\Delta_1\sqrt{\Delta_2}}
	\left(P_{11}\dd u_1+P_{12}\dd u_2+P_{13}\dd u_3\right)\,,
	\\
	Z_\rmC^\dagger\dd Z_\rmA=&\frac{1}{\sqrt{\Delta_1\Delta_2}}
	\left(P_{21}\dd u_1+P_{22}\dd u_2+P_{23}\dd u_3\right)\,,
	\\
	Z_\rmC^\dagger\dd Z_\rmB=&\frac{1}{\sqrt{\Delta_1}\Delta_2}
	\left(P_{31}\dd u_1+P_{32}\dd u_2+P_{33}\dd u_3 \right)
	\end{split}
	\label{oneform}
	\end{equation}
	where
	\begin{align}
		&\left(\begin{array}{ccc}
		P_{11}&P_{12}  &P_{13}  \\ 
		P_{21}&P_{22}  &P_{23}  \\ 
		P_{31}&P_{32}  & P_{33}
		\end{array} 
		\right)
		\notag \\
		&=
		\left(
		\begin{array}{ccc}
		1-u_1^*u_2u_3^*+|u_2|^2&-u_1u_2^*+u_3^*+u_3^*|u_1|^2  & 0 \\ 
		-u_3& 1 & 0 \\ 
		-u_3\left(u_1^*+u_2^*u_3\right)& u_1^*+u_2^*u_3 & -\Delta_1
		\end{array} \right).
		\label{formP}
	\end{align}
	
	\subsection{Topological properties}
	We consider static configurations in two space dimensions $(x^1,x^2)$ and also with wave components traveling along $x^3$ 
	axis with the speed of light. The vectors \eqref{explict Z} provide a mapping from 
	$\mathbb{R}^2$ (or $\mathbb{R}^2\times\mathbb{R}^2$) into $F_2$.
	However, for finiteness of the energy, the vectors should go to a constant at infinity on the $x^1x^2$ plane. 
	Thus, the plane can be identified as $S^2$.
	Since both the static and time-dependent configurations localize only on the plane, 
	they are topologically classified into homotopy classes of a mapping 
	$S^2\to F_2$
	, i.e. $\pi_2(F_2)$.
	In addition, there exists an useful theorem 
	(see e.g. \cite{DAdda:1978vbw}); $\pi_2(G/H)=\pi_1(H)_G$ where $\pi_1(H)_G$ is the subset 
	of $\pi_1(H)$ formed by closed paths in $H$ which can be contracted to a point in $G$. 
	Therefore, these configurations are characterized by the second homotopy group  
	\begin{equation}
	\pi_2(F_2)=\pi_2\left(SU(3)/U(1)^2\right)=\pi_1\left(U(1)^2\right)=\mathbb{Z}+\mathbb{Z}
	\label{homotopy_f2}
	\end{equation}
	which guarantees that the solutions are classified by two independent integers. 
	Note that there are three topological degrees defined in \eqref{Na}, but  
	independent degrees are just two due to the identity \eqref{constraint N}.

	We can also define a topological invariant by means of \Kahler structure of the flag space $F_2$, though $F_2$ is not \Kahler manifold.
	Note that the non-symmetric manifolds, unlike symmetric manifolds, possesses a family of invariant metrics including both \Kahler and non-\Kahler metrics. 
	In order to simplify the discussion, 
	we consider the K$\ddot{\mathrm{a}}$hler case.  
	Then the corresponding \Kahler potential can be defined as
	\begin{equation}
	K=K_1+K_2
	\end{equation}
	where $K_j=m_j\log\Delta_j$, $j=1,2$, with $m_j$ being a positive constant.   
	Since the manifold is now the \Kahler one, it possesses the closed two-form $\Omega$
	 i.e. $\dd \Omega=0$. Here, $\dd=\pd+\bar{\pd}=\dd u_\alpha\frac{\pd}{\pd u_\alpha}+\dd u_\beta^*\frac{\pd}{\pd u_\beta^*}$ denote the exterior derivative, 
	while the operator $\pd$ and $\bar{\pd}$ are  called the Dolbeault operators. 
	The condition $\dd \Omega=0$ is equivalent to 
	\begin{equation}
	\Omega=i\pd\bar{\pd}K.
	\end{equation}
	If one divides the two-form into two parts
	$
	\Omega_j=i\pd\bar{\pd}K_j 
	$
	,they are given by 
	\begin{align}
	\Omega_1=\frac{im_1}{\Delta_1^2}&\left\{
	\left(1+|u_2|^2\right)\dd u_1\wedge \dd u_1^*
	-u_2u_1^*\dd u_1\wedge \dd u_2^*
	\right.
	\notag\\
	&
	\left.\quad
	-u_1u_2^*\dd u_2\wedge \dd u_1^*
	+\left(1+|u_1|^2\right)\dd u_2\wedge \dd u_2^*
	\right\}\,, 
	\\
	\Omega_2=\frac{im_2}{\Delta_2^2}&\left\{
	\left(1+|u_3|^2\right)
	\dd\left(u_1u_3-u_2\right) \wedge \dd \left(u_1^*u_3^*-u_2^*\right)
	\right.
	\notag\\
	&\left. \quad
	-u_3\left(u_1^*u_3^*-u_2^*\right)
	\dd \left(u_1u_3-u_2\right)\wedge \dd u_3^* \right. \notag\\
	&\left.\quad	
	-u_3^*\left(u_1u_3-u_2\right)
	\dd u_3\wedge \dd \left(u_1^*u_3^*-u_2^*\right)
	\right.
	\notag\\
	&\left. \quad
	+\left(1+|u_1u_3-u_2|^2\right)\dd u_3\wedge \dd u_3^*
	\right\}\,.
	\end{align}
	The topological invariant $Q$ is defined in terms of the integral of the \Kahler two-form
	\begin{equation}
	Q\equiv Q_1+Q_2, \qquad Q_j=\int \Omega_j\,.
	\end{equation}
	The relation of the topological invariant and the winding numbers is directly calculable by the integration
	\begin{align}
		\begin{split}
		Q_1&=-im_1\int\dd^2x ~\varepsilon^{ij}\left(D_i^{(\rmA)}Z_\rmA\right)^\dagger \left(D_j^{(\rmA)}Z_\rmA\right)
		\\
		&=2\pi m_1N_\rmA\,,
		\end{split}
		\\
		\begin{split}
		Q_2&=im_2\int\dd^2x ~\varepsilon^{ij}\left(D_i^{(\rmC)}Z_\rmC\right)^\dagger \left(D_j^{(\rmC)}Z_\rmC\right)
		\\
		&=-2\pi m_2N_\rmC\,.
		\end{split}
	\end{align}
	Consequently, after an appropriate normalization, 
	i.e. $m_1=m_2=\frac{1}{2\pi}$, 
	the topological invariant is characterized just by two integers $N_\rmA$ and $N_\rmC$ as
	\begin{equation}
	Q=N_\rmA-N_\rmC
	\label{top charge}
	\end{equation}
	which of course is consistent with the formal argument \eqref{homotopy_f2}.
	
	There are exact expressions of the topological invariants corresponding to \eqref{homotopy_f2}, 
	However, the field configurations in this model are characterized by two 
	integers and it is not so straightforward to find proper form of the 
	topological charge interpreted as the net number of solitons. 
	The definition \eqref{top charge} seems suitable for the net number of solitons for the parametrization \eqref{explict Z}. 
	
	As we shall see in Sec.V, the static energy of solutions is proportional to \eqref{top charge} like standard 
	BPS type solutions and then we call (\ref{top charge}) as {\it the topological charge}.

		\begin{center}
		\begin{table}[t]
		\includegraphics[width=5.0cm,clip]{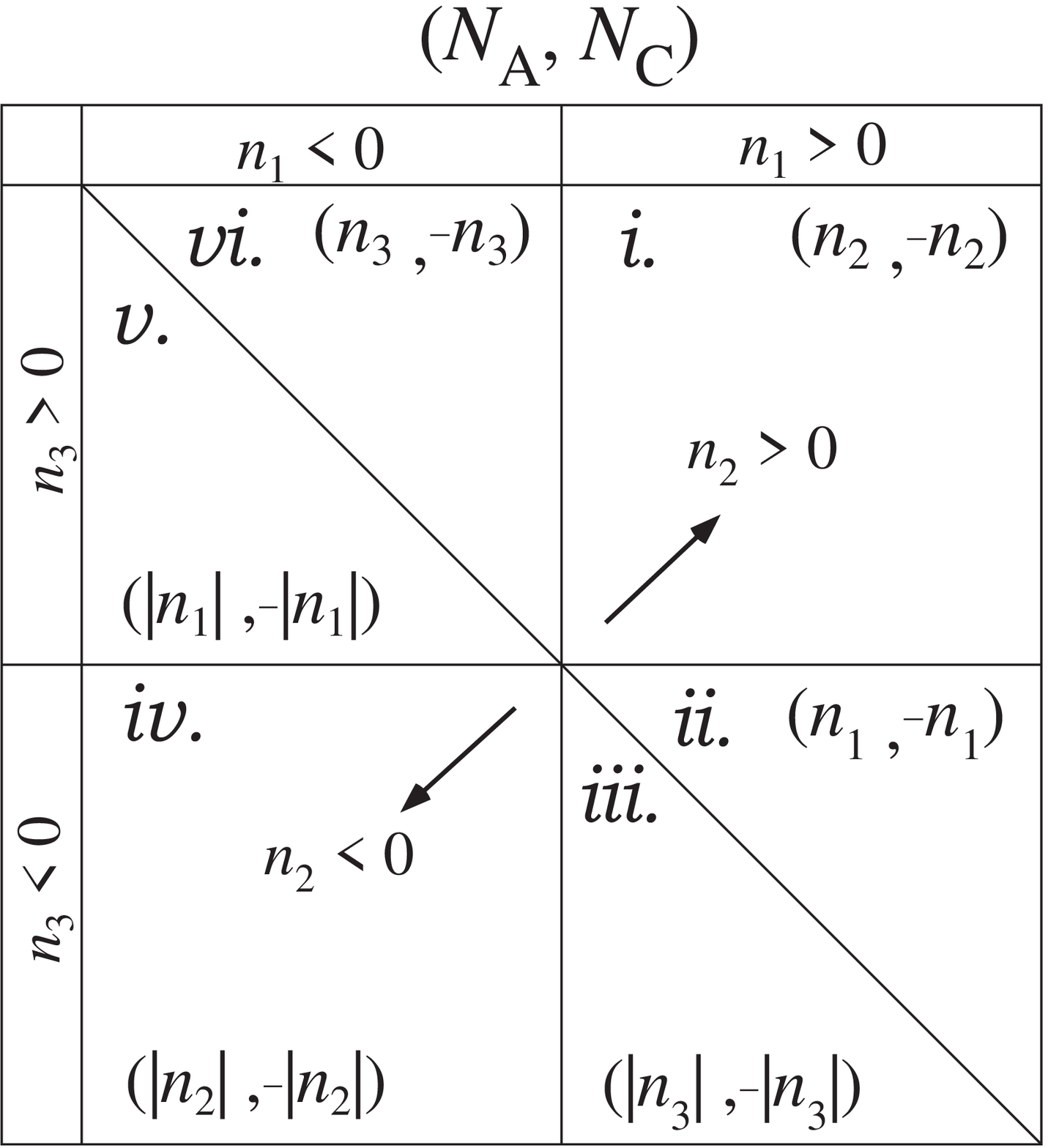}
		\caption{The values of $(N_{\rm A},N_{\rm C})$ for the winding numbers $n_1,n_2, n_3$. 
		As was shown in Sec.III-B, 
		the topological charge is given by $Q=N_\rmA-N_\rmC$. }
		\label{Tab TopCharge}
		\end{table}
		\end{center}

	\section{Solutions of the Euler-Lagrange equation}
	For finding genuine solutions, 
	we solve the Euler-Lagrange equation of the Lagrangian \eqref{F2 nls}, instead of first order coupled equations. 
	The Euler-Lagrange equation is given by
	\begin{equation}
	\partial^\mu J_\mu=0,\qquad J_\mu:=U\mathcal{B}_\mu U^\dagger
	\label{eq J}
	\end{equation}
	with
	\begin{align}
	\mathcal{B}_\mu&=\left( \begin{array}{ccc}
	0  &Z_\rmA^\dagger\partial_\mu Z_\rmB  &Z_\rmA^\dagger\partial_\mu Z_\rmC  \\ 
	Z_\rmB^\dagger\partial_\mu Z_\rmA&0 &Z_\rmB^\dagger\partial_\mu Z_\rmC  
	\\ 
	Z_\rmC^\dagger\partial_\mu Z_\rmA& Z_\rmC^\dagger\partial_\mu Z_\rmB & 0
	\end{array} \right)\,. 	
	\label{expB}
	\end{align}
	Here $J_\mu$ is the Noether current associated to the global $SU(3)$ symmetry.
	The equation \eqref{eq J} can be rewritten as
	\begin{equation}
	\pd^\mu \mathcal{B}_\mu +\left[\mathcal{A}^\mu, \mathcal{B}_\mu\right]=0
	\label{eq Lax}
	\end{equation}
	where $\mathcal{A}_\mu$ is a flat connection defined by
	\begin{equation}
	\mathcal{A}_\mu=U^\dagger\partial_\mu U\,.
	\end{equation}
	From \eqref{eq Lax} with \eqref{expB} we obtain the coupled equations 
	\begin{equation}
		\begin{split}
		&\partial^\mu
		\left(
		Z_\rmB^\dagger\partial_\mu Z_\rmA
		\right) 
		+\left(Z_\rmB^\dagger\partial^\mu Z_\rmB-Z_\rmA^\dagger\partial^\mu Z_\rmA\right)
		Z_\rmB^\dagger\partial_\mu Z_\rmA
		=0 ,
		\\
		&\partial^\mu
		\left(
		Z_\rmC^\dagger\partial_\mu Z_\rmA
		\right) 
		+\left(Z_\rmC^\dagger\partial^\mu Z_\rmC-Z_\rmA^\dagger\partial^\mu Z_\rmA\right)
		Z_\rmC^\dagger\partial_\mu Z_\rmA
		=0 ,
		\\
		&\partial^\mu
		\left(
		Z_\rmC^\dagger\partial_\mu Z_\rmB
		\right) 
		+\left(Z_\rmC^\dagger\partial^\mu Z_\rmC-Z_\rmB^\dagger\partial^\mu Z_\rmB\right)
		Z_\rmC^\dagger\partial_\mu Z_\rmB
		=0 ,
	\end{split}
	\label{eq Z}
	\end{equation}
	and also their complex conjugations. 
	Plugging the explicit form of the vectors 
	\eqref{explict Z}, \eqref{formdelta} and the one-form {\eqref{oneform} into the equations, 
	we finally obtain the equations for the scalar fields
	\begin{widetext} 	
	\begin{align}
	&\DOne\DTwo\left(P_{11}\pd^\mu\pd_\mu u_1+P_{12}\pd^\mu\pd_\mu u_2\right)
	-2\DTwo\left(P_{11}\pd_\mu u_1+P_{12}\pd_\mu u_2\right)
	\left(u_1^*\pd^\mu u_1+u_2^*\pd^\mu u_2\right) \notag\\
	&\hspace{1cm}-\DOne
	\left[\left(u_1+u_2u_3^*\right)\left(u_3^*\pd^\mu u_1^*-\pd^\mu u_2^*\right)+\DOne\pd^\mu u_3^*\right]
	\left(u_3\pd^\mu u_1-\pd^\mu u_2\right)=0,
	\label{eq BA u} 
	\\
	&\DOne\DTwo\left(P_{21}\pd^\mu\pd_\mu u_1+P_{22}\pd^\mu\pd_\mu u_2-\pd^\mu u_3\pd_\mu u_1\right)
	  \notag\\
	&\hspace{1cm}-\left(P_{21}\pd_\mu u_1+P_{22}\pd_\mu u_2\right)
	\left[\DTwo\left\{u_1^*\pd^\mu u_1+u_2^*\pd^\mu u_2 \right\}+
	\DOne\left\{\left(u_1^*u_3^*-u_2^*\right)\pd^\mu\left(u_1u_3-u_2\right) +u_3^*\pd^\mu u_3 \right\}\right]=0,
	\label{eq CA u} \\
	&\DOne\DTwo\left\{P_{31}\pd^\mu\pd_\mu u_1+P_{32}\pd^\mu\pd_\mu u_2+P_{33}\pd^\mu\pd_\mu u_3-2\left(u_1^*+u_2^*u_3\right)\pd_\mu u_1\pd^\mu u_3\right\}
	\notag\\
	&\hspace{1cm}-2\DOne\left(P_{31}\pd_\mu u_1+P_{32}\pd_\mu u_2+P_{33}\pd_\mu u_3\right)
	\left\{\left(u_1^*u_3^*-u_2^*\right)\pd^\mu\left(u_1u_3-u_2\right) +u_3^*\pd^\mu u_3 \right\}  \notag\\
	&\hspace{1cm}-\DTwo
	\left[\left(1-u_1u_2^*u_3+|u_2|^2\right)\pd_\mu u_1^*+
	\left(-u_1^*u_2+u_3+u_3|u_1|^2\right)\pd_\mu u_2\right]
	\left(u_3\pd^\mu u_1-\pd^\mu u_2\right)=0,
	\label{eq CB u}
	\end{align}
	\end{widetext}
	where the explicit form of $\Delta_1, \Delta_2$ and $P_{ij}~i,j=1,2,3$ are given in 
	\eqref{formdelta} and \eqref{formP}, respectively.


	\begin{figure*}[t!]
		\begin{minipage}[b]{0.32\linewidth}
			\centering
			\includegraphics[width=5.5cm,clip]{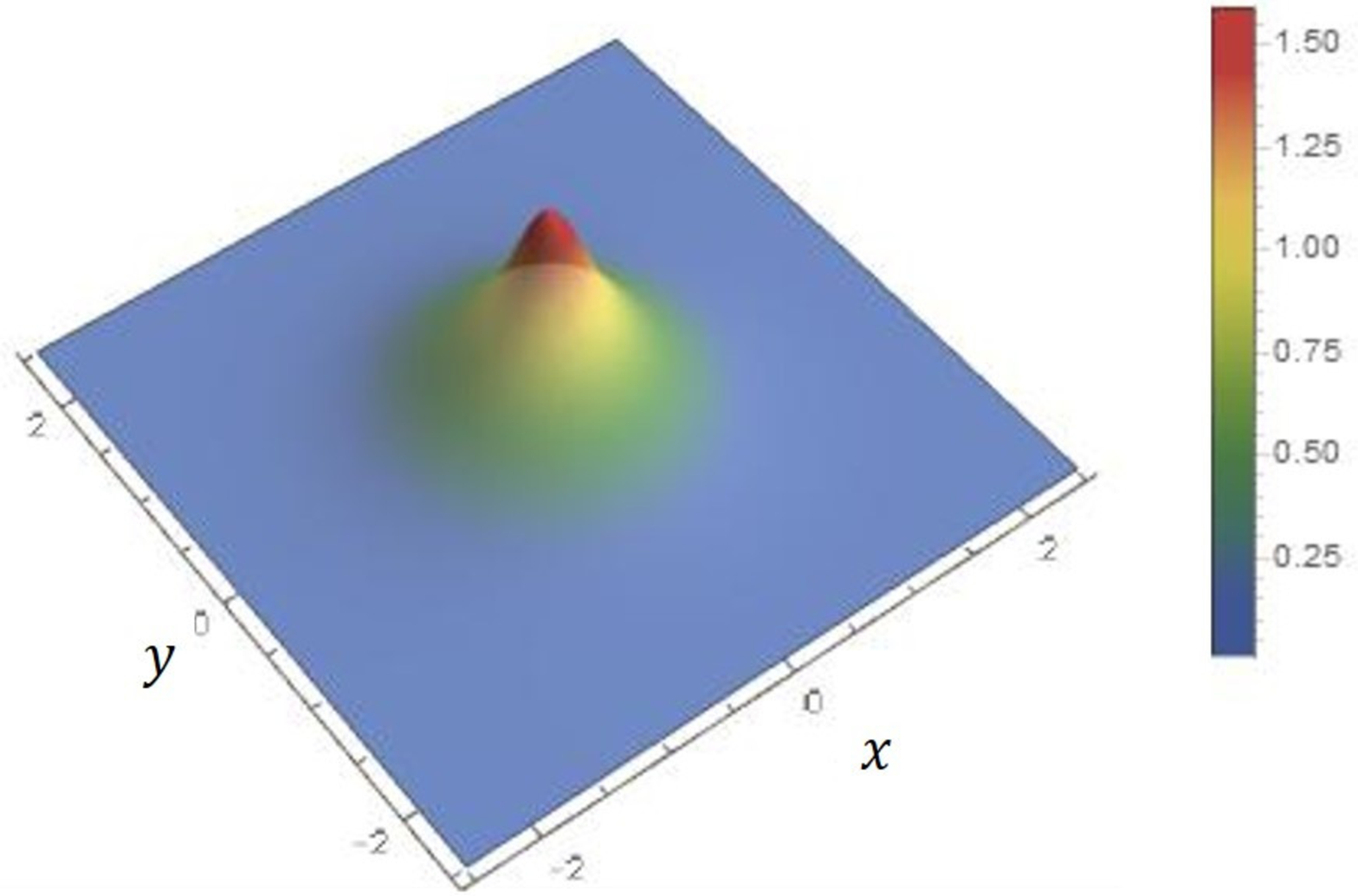}
		\end{minipage}
		\begin{minipage}[b]{0.32\linewidth}
			\centering
			\includegraphics[width=5.5cm,clip]{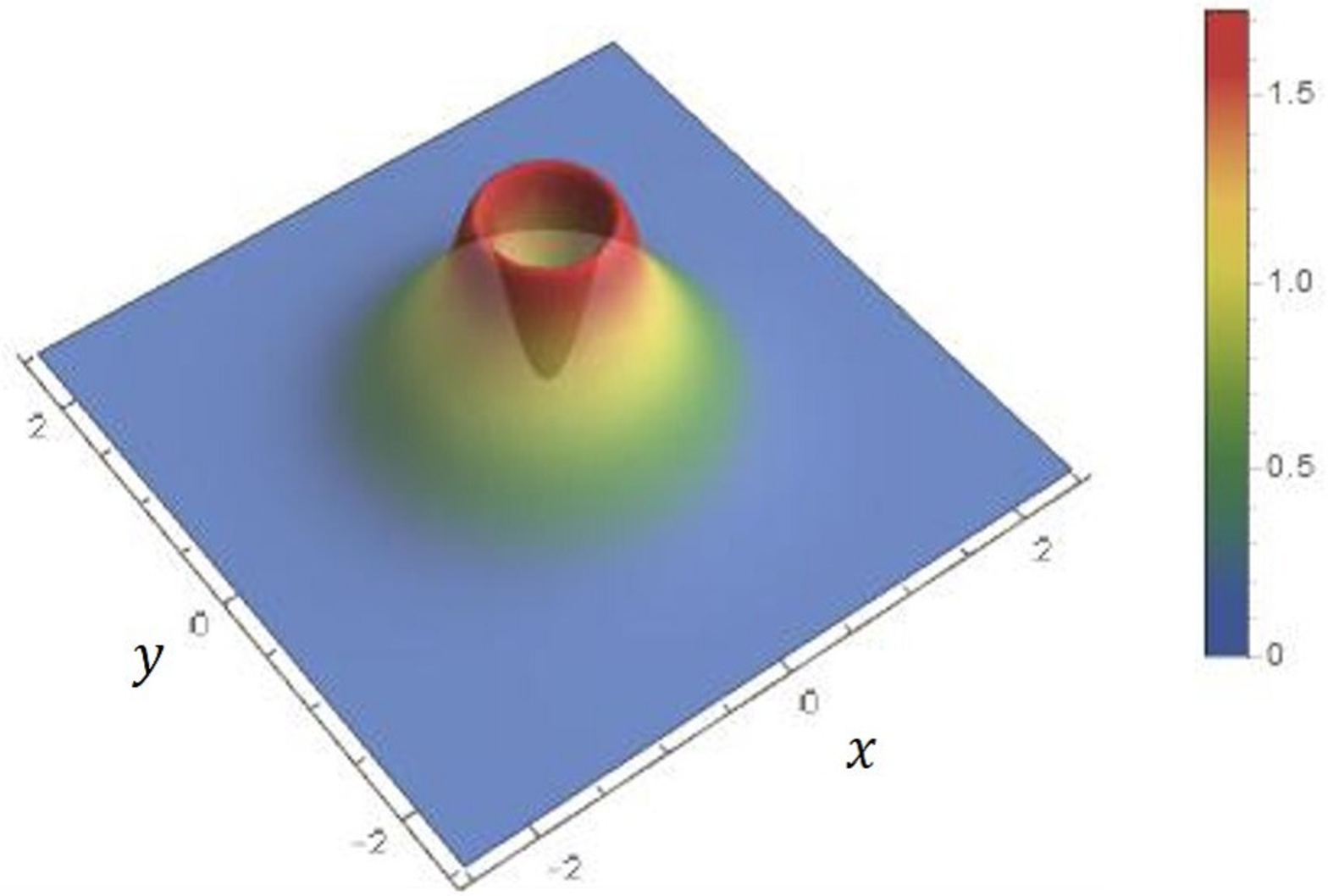}
		\end{minipage}
		\begin{minipage}[b]{0.32\linewidth}
			\centering
			\includegraphics[width=5.5cm,clip]{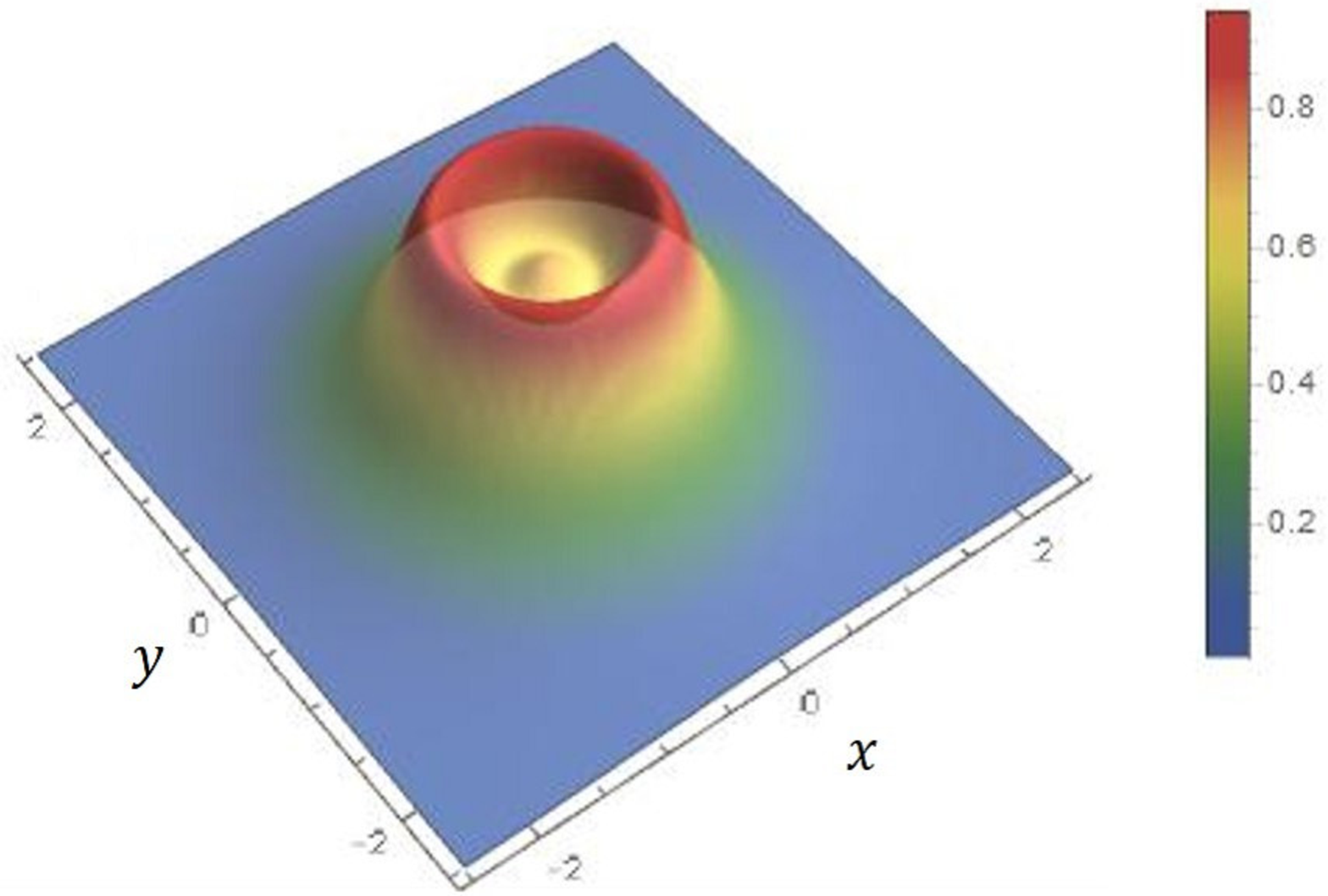}
		\end{minipage}
		\caption{The topological charge density for the solution
			$(u_1,u_2,u_3)=(z^{n_1}, z^{n_2}, \frac{n_2}{n_1}z^{n_2-n_1})$ with the winding numbers
			$(n_1,n_2)=(1,2), (1,3), (2,3)$, from left to right.}
			\label{FIG Topcharge}
	\end{figure*}
	


	\begin{figure*}[t!]
		\begin{minipage}[b]{0.32\linewidth}
			\centering
			\includegraphics[width=5.5cm,clip]{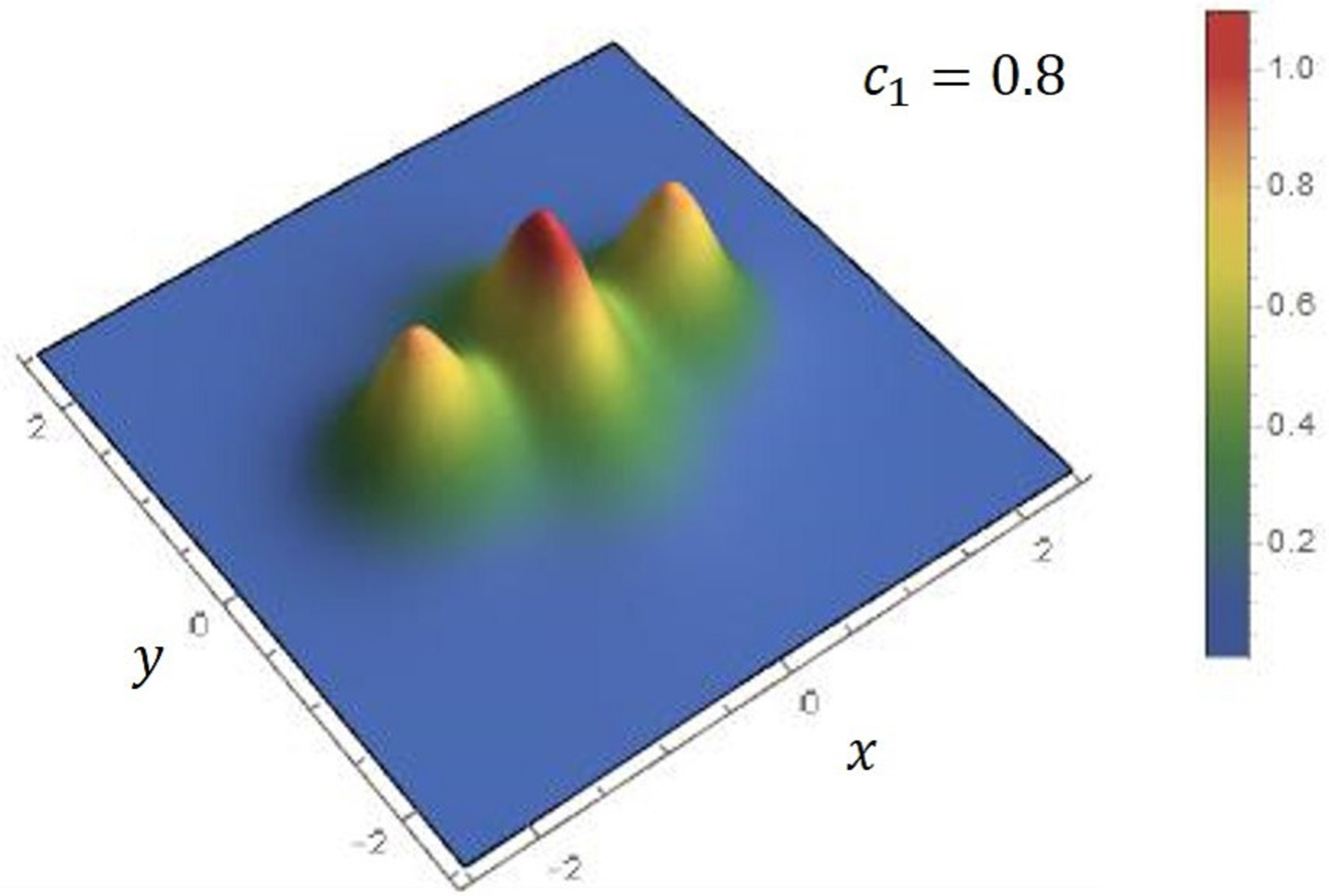}
		\end{minipage}
		\begin{minipage}[b]{0.32\linewidth}
			\centering
			\includegraphics[width=5.5cm,clip]{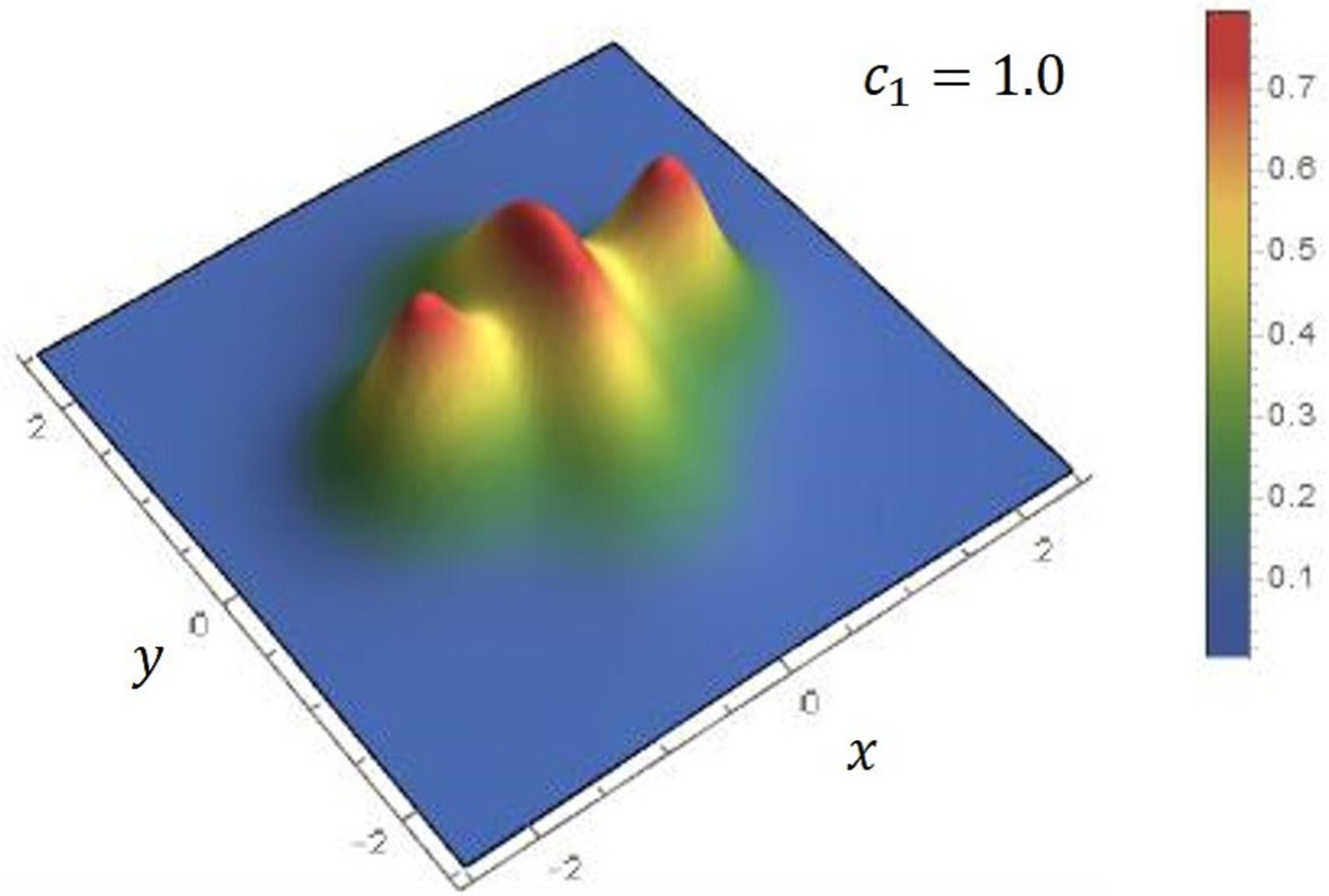}
		\end{minipage}
		\begin{minipage}[b]{0.32\linewidth}
			\centering
			\includegraphics[width=5.5cm,clip]{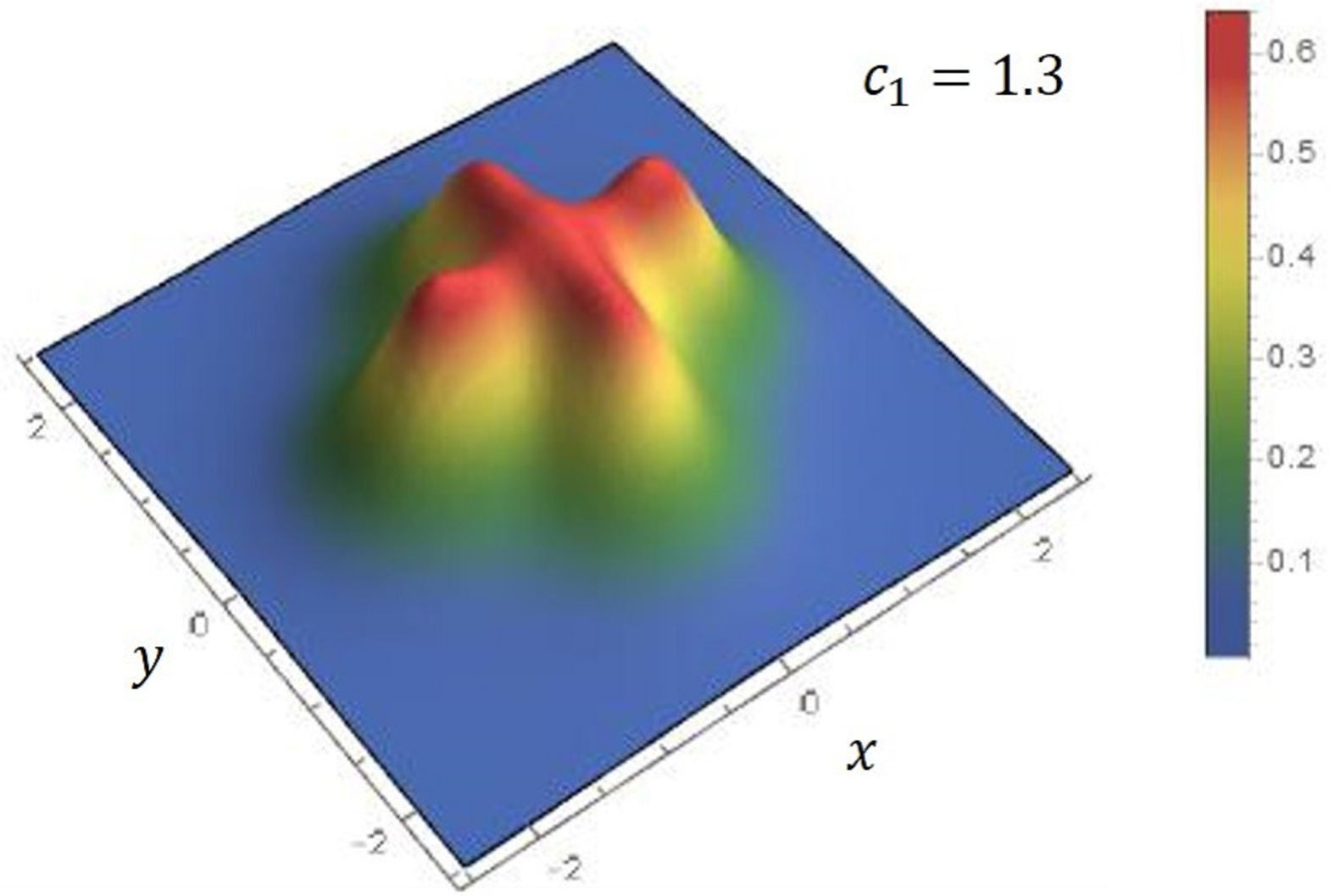}
		\end{minipage}\\
		\begin{minipage}[b]{0.32\linewidth}
			\centering
			\includegraphics[width=5.5cm,clip]{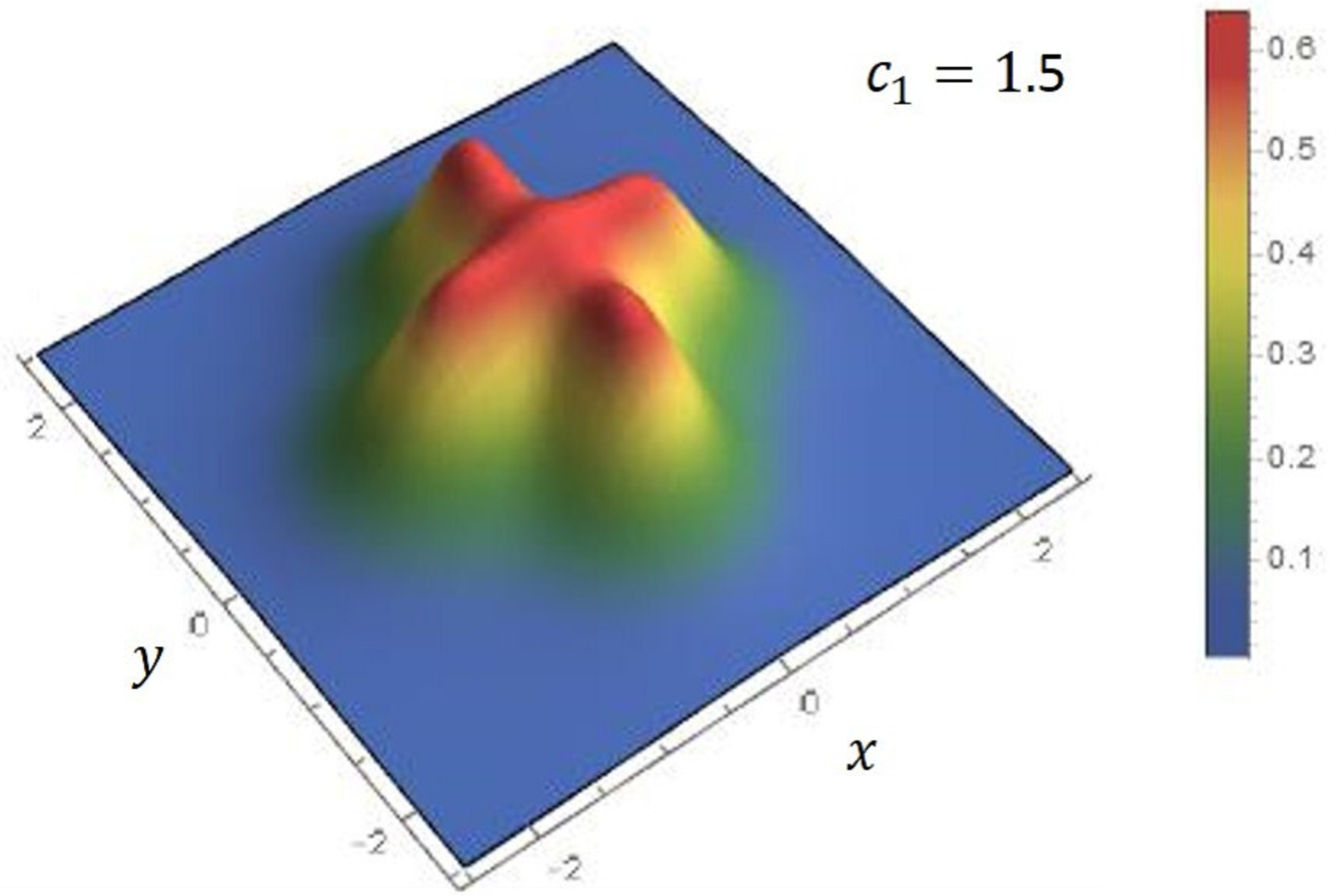}
		\end{minipage}
		\begin{minipage}[b]{0.32\linewidth}
			\centering
			\includegraphics[width=5.5cm,clip]{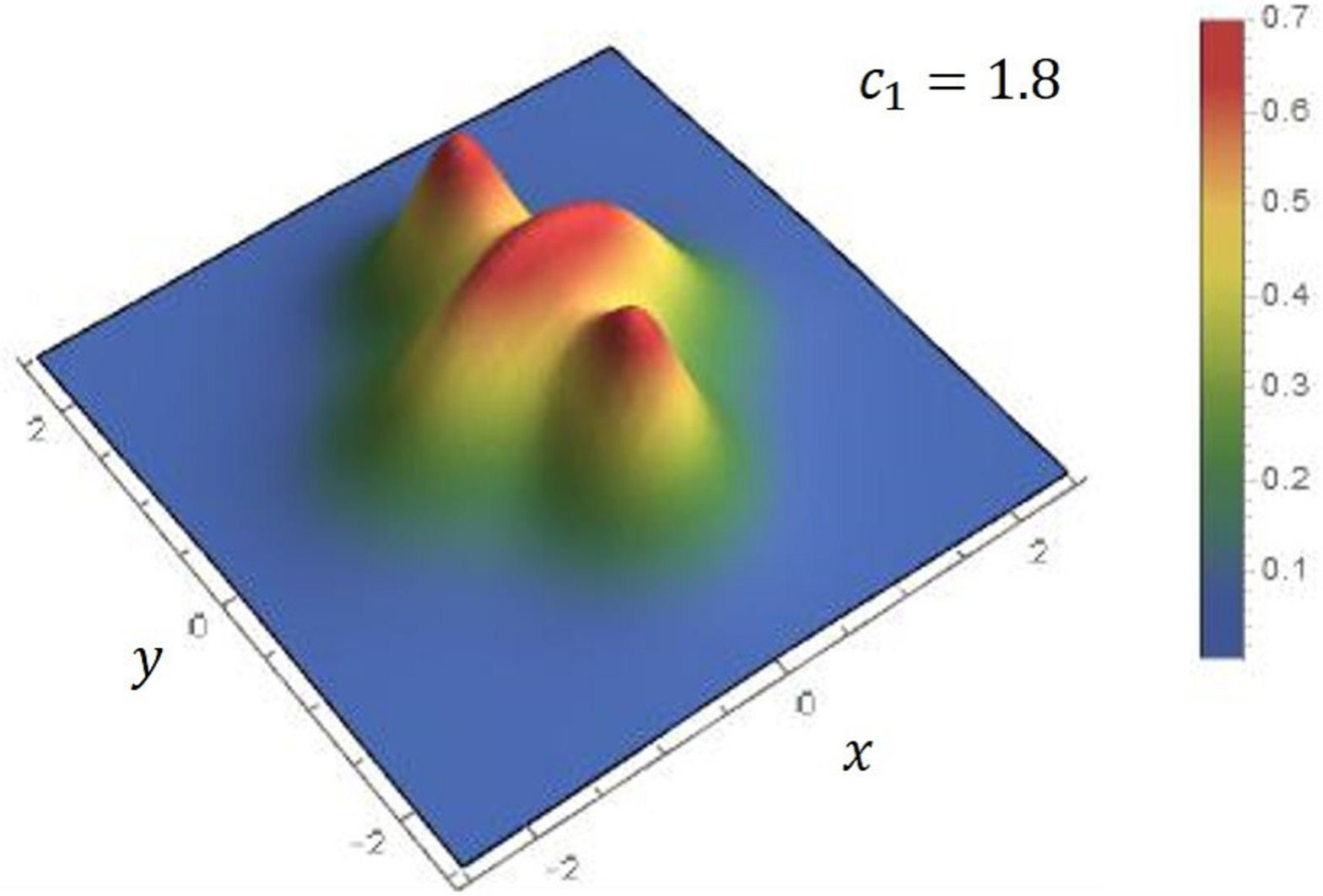}
		\end{minipage}
		\begin{minipage}[b]{0.32\linewidth}
			\centering
			\includegraphics[width=5.5cm,clip]{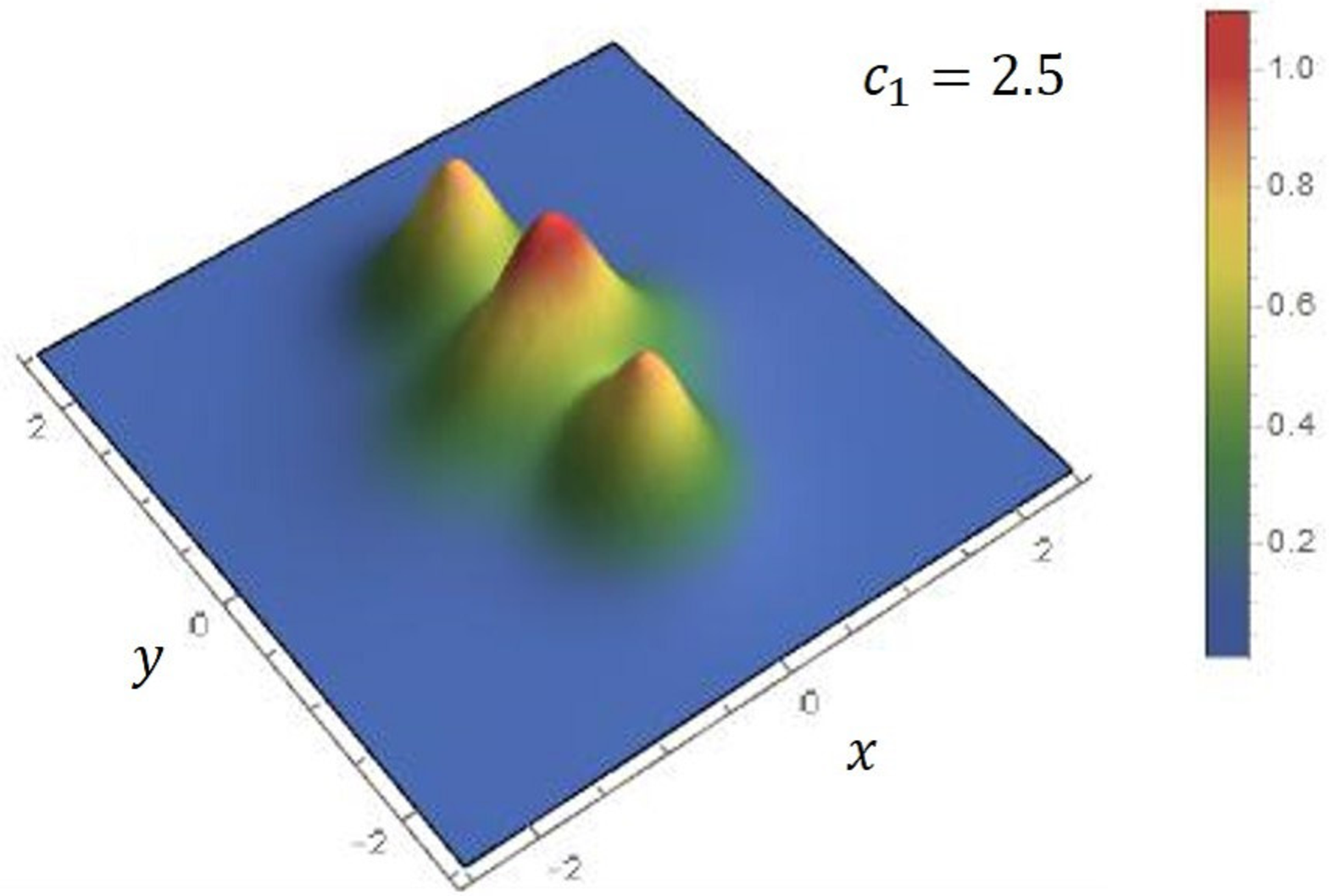}
		\end{minipage}
		\caption{The topological charge density for 
			$(u_1,u_2,u_3)=(c_1 z,z^2-1,2z/c_1)$ with the parameter 
		$c_1=0.8, 1.0, 1.3, 1.5, 1.8, 2.5$, from upper left to lower right.  }
		\label{FIG Topcharge_aniso}
	\end{figure*}

	Since the equations are quite complicated and highly non-linear, 
	for the moment we introduce the simplified ansatz; all the complex scalars $u_i$ depend only on $z$ and $y_+$
	($y_\pm$ is defined as $y_\pm=x^0\pm x^3$), i.e.
	\begin{equation}
		u_i=u_i(z,y_+) \quad\text{and}\quad u_i^*=u_i^*(\bar{z},y_+) ~. \label{ansatz}
	\end{equation}
	It is easy to see that the ansatz satisfies
	\begin{equation}
		\pd^\mu\pd_\mu u_i=0~,\qquad
		\pd^\mu u_i\pd_\mu u_j=0 ~.
	\end{equation}
	One can directly check that the second equation \eqref{eq CA u} is automatically satisfied by the ansatz \eqref{ansatz}. 
	\eqref{eq BA u} and \eqref{eq CB u} are certainly simplified
	\begin{align}
		&\left[P_{32}^*\left(u_3^*\pd^\mu u_1^*-\pd^\mu u_2^*\right)+\DOne\pd^\mu u_3^*\right]
		\left(u_3\pd^\mu u_1-\pd^\mu u_2\right)=0,
		\label{eq hol1}
		\\
		&\left[P_{11}^*\pd^\mu u_1^*+
		P_{12}^*\pd^\mu u_2^*\right]
		\left(u_3\pd^\mu u_1-\pd^\mu u_2\right)=0.
		\label{eq hol2}
	\end{align}
	Therefore, if we find the field configurations satisfying
	\begin{equation}
		u_3\pd_\mu u_1-\pd_\mu u_2=0,
		\label{Key equation}
	\end{equation}
	they exactly solve \eqref{eq BA u} - \eqref{eq CB u}. 

	First we consider the static solutions within the ansatz (\ref{ansatz}) which satisfy (\ref{Key equation}).
	Note that if one scalar field is set to a constant, there are only trivial and embedding solutions.
	In the intriguing case where all the scalars aren't constants, the solutions
	can generally be written as
	\begin{equation}
	u_1=\frac{p_1(z)}{q_1(z)}, \qquad u_2=\frac{p_2(z)}{q_2(z)}, \qquad
	u_3=\frac{\pd_z u_2}{\pd_z u_1}
	\label{static sol}
	\end{equation}
	where $p_i(z)$ and $q_i(z)$ are irreducible polynomials of $z$ and $u_1$ isn't proportional to $u_2$. 
	The winding number of the scalar field $u_i$ is equal to number of the poles of the scalar field including those at infinity
	and thereby
	\begin{equation}
	\begin{split} 
	&n_i=\mathrm{max} \{\mathrm{deg}(p_i),\mathrm{deg}(q_i)\},\quad i=1,2 
	 \\
	&n_3=n_2-n_1.
	\end{split}
	\end{equation}
	According to the derivation of the topological charge in the $CP^N$ model \cite{DAdda:1978vbw}, 
	the topological degrees $N_{\rm A}$ and $N_{\rm C}$ are given 
	by pair of the winding numbers $n_i$ as 
	\begin{align}
	&N_{\rm A}=\mathrm{max}( 0, n_1, n_2)-\mathrm{min}(0, n_1, n_2)\,, 
	\\
	&N_{\rm C}=\mathrm{max}( -n_2,-n_3, 0)-\mathrm{min}(-n_2,-n_3, 0)\,.
	\end{align}
	As we summarize in Table \ref{Tab TopCharge}, the topological degrees $(N_{\rm A},N_{\rm C})$ 
	are thus defined by pair of the winding numbers with opposite signs.

	\begin{center}
		\begin{figure*}[t]
			\includegraphics[width=18cm,clip]{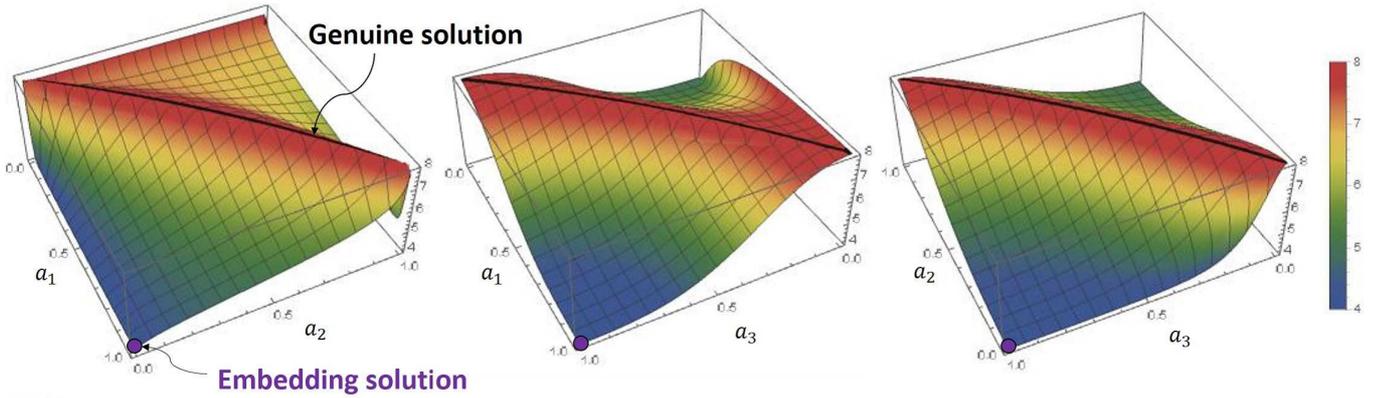}
			\caption{Energy surface for $u_1=\frac{1-a_1}{a_1}z^3,u_2=\frac{1-a_2}{a_2}z^4,u_3=\frac{1-a_3}{a_3}z$.
				From left to right $a_3=0.5$,  $a_2=0.5$, $a_1=0.5$}
			\label{surface}
		\end{figure*}
	\end{center}
	

	The most simple choice of the static solution \eqref{static sol} seems 
	\begin{equation}
	u_1=c_1z^{n_1},~~ u_2=c_2z^{n_2} ,~~ u_3=\frac{c_2n_2}{c_1n_1}z^{n_2-n_1} 
	\label{hol sol}
	\end{equation}
	where $c_i$ are non-zero complex constants. 
	In Fig.\ref{FIG Topcharge}, we plot the topological charge density of \eqref{hol sol}. 
	In the case of $(n_1,n_2)=(1,2)$, it is lump shaped of which the peak locates at the origin, 
	while for higher winding numbers, the solutions always exhibit the crater like structure.
	Anisotropic configuration can also be examined. The solution   
	\begin{align}
	u_1=c_1 z,~~u_2=z^2-1,~~u_3=\frac{2}{c_1}z
	\end{align}
	exhibits non circular shape of the density (see Fig.\ref{FIG Topcharge_aniso}). 
	For growing the parameter $c_1$, the lumps collide and scatter with the right angle.

	Following to the prescription for construction of the $CP^N$ vortices ~\cite{Ferreira:2008nn,Ferreira:2010jb,Ferreira:2011nd}, 
	we examine a class of time-dependent solutions which is made by a product of localizing component on the $x^1x^2$ plane
	and traveling wave being parallel to $x^3$ with speed of light, i.e. ansatz of the form $u_i=f_i(z)w_i(y_{+}),~i=1,2,3$.
	Again, the nonembedding time-dependent solutions are not allowed for the case that one or more of the functions $f_i$ and $w_i$ remain constant.  
	When all $f_i$ and $w_i$ are the functions,   
	by substituting the ansatz into \eqref{Key equation} one easily obtain the conditions
	\begin{align}
		&f_2=\alpha f_1^C, \qquad f_3=\alpha A f_1^{C-1}\,,
		\\
		&w_2=\beta w_1^C, \qquad w_3=\beta B w_1^{C-1}
	\end{align}
	where $\alpha, \beta, A$ and $B$ are non-zero constants, and $C=A/B$.
	One of the solutions which keep single-valuedness and finiteness of the energy per unit length is given as
	\begin{align}
	 u_1&=c_1z^{n_1}e^{in_1ky_{+}} ~,
	 \notag\\
	 u_2&=c_2z^{n_2}e^{in_2ky_{+}} ~,
	\label{wave sol}
	\\
	 u_3&=\frac{c_2n_2}{c_1n_1}z^{n_2-n_1} e^{i(n_2-n_1)ky_{+}} \notag
	\end{align}
	where $k$ is a constant being proportional to the inverse of a wavelength.

	Note that under the constraint \eqref{Key equation}, there is a sure relation among the vectors $Z_a$; $D^{(\rmA)}_\mu Z_\rmA$ 
	is proportional to  $Z_\rmB$ and similarly $D^{(\rmB)}_\mu Z_\rmB$ is to $Z_\rmC$. 
	Consequently, the vectors correspond to the tower of the general solution in the $CP^2$ model generated through the \Backlund transformation \cite{Din:1980jg,Ferreira:2011jy}.
	It implies that for the solution \eqref{static sol} and \eqref{wave sol}, $Z_\rmA$ can be recognized as an instanton, $Z_\rmC$ as an anti-instanton, and $Z_\rmB$ as an instanton-anti instanton bound state.	
	Similar types of solutions are already discussed in ~\cite{Bykov:2015pka}, 
	but the model the author employed is different from ours because it contains a non-zero Kalb-Ramond field which contributes to the Euler-Lagrange equation. 	
	We have shown that the extra term has no effect at least for existence of the analytical solutions.
	Further, as far as we know that any time-dependent solutions of the $F_2$ model have not been discussed previously.

\begin{center}
	\begin{figure*}[t]
		\includegraphics[width=18cm,clip]{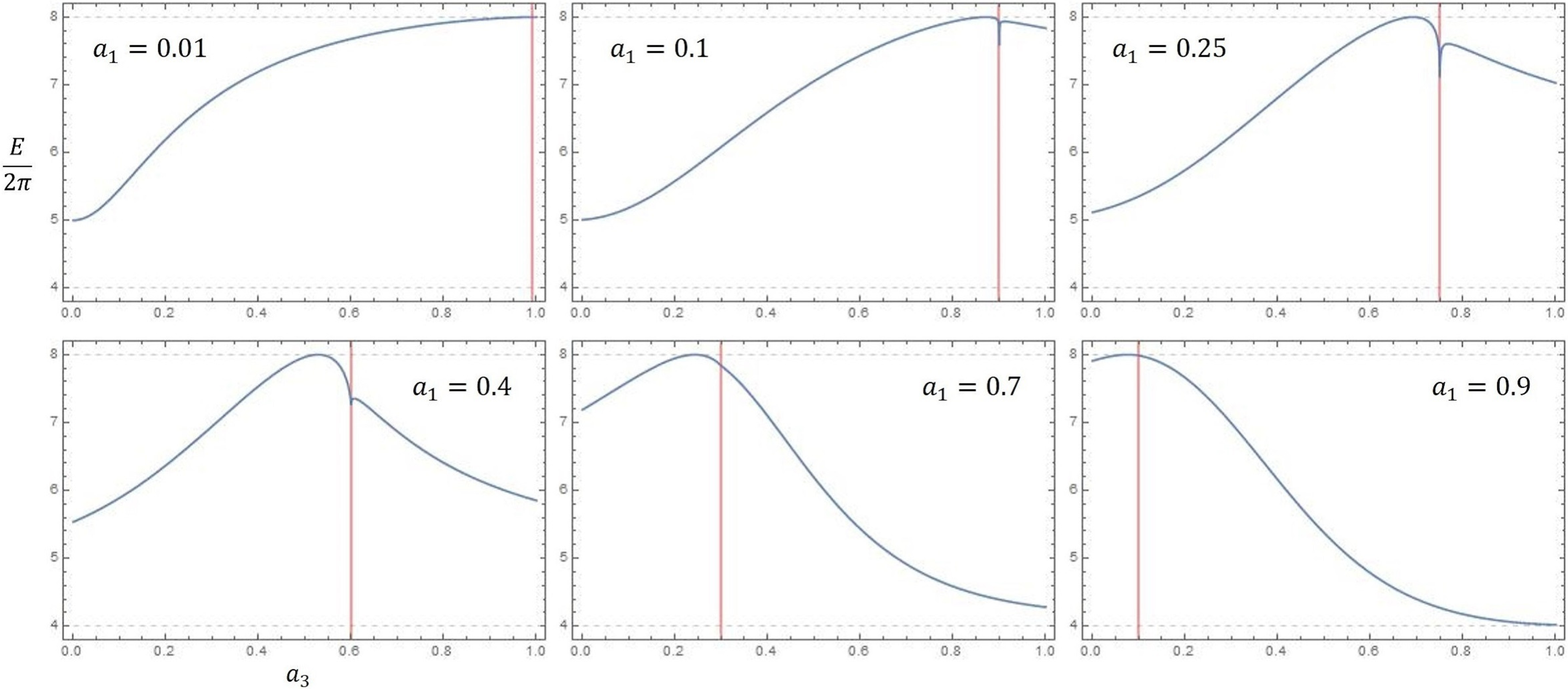}
		\caption{\label{energy} The cuts of the total energy (per $2\pi$) corresponding to Fig.\ref{surface} with $a_2=0.5$.
			The red lines denote a barrier satisfying $u_1u_3=u_2$ which is prohibited by the topological reason. 
			There are always dips on the barrier and maxima near the line, but both of them are not stationary points. }
	\end{figure*}
\end{center}

	\section{Property of The energy}

	In this section, we discuss the energy of the static solutions \eqref{static sol} and also of the wave solution \eqref{wave sol}. 
	We show that the static solutions saturate to an energy lower bound different from \eqref{Bound 1}. 
	It is proved that the solutions \eqref{static sol} are obtained through coupled first derivative equations 
	which correspond to the saturation condition. 
	For the traveling wave solution \eqref{wave sol}, the energy can be written by the topological charge plus
	Noether charges which are relevant to some $U(1)$ symmetries.
	
	Using the completeness condition \eqref{completeness}, 
	the static energy per unit length \eqref{Energy1} can be written as
	\begin{align}
	E=\int \dd^2x
	\left(\left\|D^{(\rmA)}_i Z_\rmA\right\|^2
	+\left\|D^{(\rmC)}_i Z_\rmC\right\|^2
	-\left\|Z_\rmC^\dagger\partial_i Z_\rmA\right\|^2\right).
	\label{energy2}
	\end{align}
	We again emphasize that the first and second term in \eqref{energy2} are the static energy of the $CP^2$ model.
	Applying the BPS bound of the $CP^2$ model to the two terms, 
	we obtain the energy bound	
	\begin{align}
	E=&
	\frac{1}{2}\int \dd^2x
	\left\|D^{(\rmA)}_i Z_\rmA\mp i\varepsilon^{ij}D^{(\rmA)}_j Z_\rmA\right\|^2
	 \notag\\
	+&\frac{1}{2}\int \dd^2x 
	\left\|D^{(\rmC)}_i Z_\rmC\mp i\varepsilon^{ij}D^{(\rmC)}_j Z_\rmC\right\|^2
	\notag\\
	\pm& i\int \dd^2x~\varepsilon^{ij}\left\{
	\left(D^{(\rmA)}_i Z_\rmA\right)^\dagger D^{(\rmA)}_j Z_\rmA
	+\left(D^{(\rmC)}_i Z_\rmC\right)^\dagger D^{(\rmC)}_j Z_\rmC
	\right\}
	\notag\\
	-&\int \dd^2x\left\|Z_\rmC^\dagger\partial_i Z_\rmA\right\|^2
	\notag\\
	\geq& 2\pi \left(|N_\rmA|+|N_\rmC|\right)
	-\int \dd^2x\left\|Z_\rmC^\dagger\partial_i Z_\rmA\right\|^2\,.
	\label{Bound 2}
	\end{align}
	According to the discussion in the $CP^2$ model, the equality in \eqref{Bound 2} is satisfied 
	when the scalar fields are functions of only $z$ (or $\bar{z}$), i.e. $u_i=u_i(z)$. 
	Thus the static solutions \eqref{static sol} obviously saturate the new bound~\eqref{Bound 2}.
	Unlike standard BPS bound, \eqref{Bound 2} has non-topological term (the 
	last term in \eqref{Bound 2}).  
	Since the maximum of the non-topological term (without minus sign) gives the equivalence between the two bounds \eqref{Bound 1} and \eqref{Bound 2}, 
	the value of the term runs 
	\begin{equation}
		0\leq\int \dd^2x\left\|Z_\rmC^\dagger\partial_i Z_\rmA\right\|^2
		\leq\pi \left(|N_\rmA|-|N_\rmB|+|N_\rmC|\right) .
		\label{Bound nontop}
	\end{equation}
	One can find from \eqref{oneform}  that $Z^\dagger_\rmC\pd_i Z_\rmA$ is proportional to $u_3\pd_i u_1-\pd_i u_2$. Therefore, for the solutions of \eqref{Key equation}, this non-topological term identically vanishes (then
	takes the lower bound of the inequality of (\ref{Bound nontop})).
	Consequently, the static energy of the nonembedding solutions
	\eqref{static sol} is given by
	\begin{equation}
		E=2\pi \left(|N_\rmA|+|N_\rmC|\right)=2\pi Q.
	\end{equation}
	On the other hand, for the embedding solutions which $u_1$ and $u_3$ are constants, the term satisfies the lower bound of the 
	inequality (\ref{Bound nontop}) and then it corresponds to the energy  bound \eqref{Bound 1} with $N_B=0$.
	Thus, we find that 
	the static energy is twice greater than that of the embedding solution 
	which belong to the same topological sector.
	This is consistent with the interpretation that the nonembedding solutions include instanton-anti instanton bound state which described by the harmonic map solutions in the $CP^2$ model.
		
	Note that, from the saturation conditions of the energy bound \eqref{Bound 2} and the lower inequality in \eqref{Bound nontop}, we obtain the following coupled first-order equations 
	corresponding to the solutions \eqref{static sol} :
	 \begin{equation}
	 	\begin{split}
	 	&D_i^{(\rmA)}Z_\rmA=i\varepsilon^{ij}D^{(\rmA)}_jZ_\rmA ,
	 	\\
	 	&D_i^{(\rmC)}Z_\rmC=-i\varepsilon^{ij}D^{(\rmC)}_jZ_\rmC ,
	 	\end{split}
	 	\label{quasi BPS1}
	 \end{equation}
	 and
	 \begin{equation}
	 	Z_\rmC^\dagger\partial_i Z_\rmA =0 .\label{quasi BPS2}
	 \end{equation}
	The equations \eqref{quasi BPS1} are equivalent to the Cauchy-Riemann equations for the complex scalar fields $u_i$ 
	and \eqref{quasi BPS2} exactly is \eqref{Key equation}.
	This clearly means that the solutions \eqref{static sol} of the Euler-Lagrange equations \eqref{eq Z} 
	can be obtained by the first-order equations \eqref{quasi BPS1} and \eqref{quasi BPS2}.
	Note that even when the saturation of the energy bound \eqref{Bound 2} is attained, the energy does not take a 
	stationary point due to the presence of the non topological term.
	According to the above discussion, however, if both \eqref{quasi BPS1} and \eqref{quasi BPS2} are satisfied, 
	the energy is stationary, i.e. maximal	energy in the bound \eqref{Bound 2}. 
	Therefore we conclude that the solutions \eqref{static sol}, which satisfy the both saturation conditions, are saddle-point.

	The solutions \eqref{static sol} possess basic properties as BPS solutions, 
	such that they saturate an energy lower bound and are the solutions of first order equations.
    They seem to be unstable, because they are not of the global minima in a given topological sector, which is fulfilled by the embedding solutions. 
	It is worth to check behavior of the energy for changing the parameters of the configuration 
	and confirm the stability nature of our solutions. 
	In Fig.\ref{surface} we plot the energy per unit length for the configuration
	\begin{align}
	u_1=\frac{1-a_1}{a_1}z^3,u_2=\frac{1-a_2}{a_2}z^4,u_3=\frac{1-a_3}{a_3}z
	\label{test conf}
	\end{align}
	with the parameters $a_i\in [0,1]$. 
	The nonembedding solutions locates on the ridge line of the surface, 
	while the embedding solutions are on the left front corner.
	There is a zero mode along the ridge, but they essentially be unstable. 
	A natural question occurs where the nonembedding solutions might eventually decay into the embedding ones. 
	In order to see this in detail, we plot the cuts of the energy surface at several values of $a_1$ for fixed $a_2=0.5$ 
	in Fig.\ref{energy}.
	For the configuration \eqref{test conf}, unless $u_1u_3=u_2$ or $u_2=0$,
	at infinities of the $x^1x^2$ plane the fields $Z_a$ approach the value
	\begin{equation}
		Z_\rmA\to\left( \begin{array}{c}
		0 \\ 
		0 \\ 
		1
		\end{array} \right) ,
		\quad
		Z_\rmB\to\left( \begin{array}{c}
		0 \\ 
		1 \\ 
		0
		\end{array} \right) ,
		\quad
		Z_\rmC\to\left( \begin{array}{c}
		1 \\ 
		0 \\ 
		0
		\end{array} \right)
		\label{boundary}
	\end{equation} 
	up to phase factors.
	The configuration \eqref{test conf} with $u_1u_3=u_2$ or $u_2=0$ 
	do not support the boundary condition \eqref{boundary} and then, they belong to a different topological sector from the nonembedding and embedding solutions.
	It clearly indicates that the solutions cannot continuously deform into the configuration which satisfy $u_1u_3=u_2$ or $u_2=0$ 
	with keeping finiteness of the energy.
	One can find the topological barrier associated with $u_1u_3=u_2$ lies between the nonembedding and the embedding solutions.
	It is a common feature among all topological classification in Table \ref{Tab TopCharge} that such a barrier 
	lies between nonembedding and embedding solutions.
	For configurations corresponding to the domain $\bm{i.}$ or $\bm{ii.}$ in the Table, as the above example, 
	the barrier exists on the line corresponding to $u_1u_3=u_2$ which do not obey the boundary condition of the solutions at infinity.
	For the domain $\bm{iv.}$ or $\bm{v.}$,  
	the condition $u_1u_3=u_2$ does not satisfy the boundary condition of solutions at the origin. 
	In this case, one can also easily prove the condition $u_1u_3=u_2$ lies between nonembedding and embedding solutions.	
	The situation for the domain $\bm{iii.}$ or $\bm{vi.}$ are little more complicate than for the others.
	Nonembedding solutions in the domain exist in region $c_1c_2c_3<0$ and, on the other hand, 
	embedding solutions are derived as
	$c_1,c_2\to\infty$ with keeping the condition $c_1=c_2c_3$, which gives $c_1c_2c_3>0$. 
	Therefore during deformation from a nonembedding solution to an embedding solution, at least one of the coefficient have to change the sign.
	At the point $c_i=0$, $i=1,2,3$, the boundary condition of the solutions are broken and therefore the barrier appears between solutions.	
	This observation indicates the nonembedding solution cannot decay into the embeddings in classical level, at least, 
	through a deformation described by variations of the parameter in \eqref{test conf}. 
	However, quantum mechanically the tunneling between these solutions may occurs and then the decay process may be able to achieve.
	Some readers may expect the other possibility.
	One can find lower energy region behind the ridges in Fig.4, but it is a fake effect from fixing one of the parameters that the ridges divide the region from an embedding solution.
	If we vary all three parameters, all configuration may finally decay into the embedding solutions through the tunneling effect. 
	   
	Next we consider the energy of the traveling wave solution \eqref{wave sol}.  From simple calculation, 
	one can get the energy as 
	\begin{align}
		E_{\text{wave}}=2\pi Q
		+8\pi &k^2 \left\{\mathcal{I}(n_1,n_2,c_1,c_2)
		\right.
		\notag
		\\
		+&\left.
		\mathcal{I}(n_2-n_1,n_2,\frac{c_2n_2}{c_1n_1},\frac{c_2(n_2-n_1)}{c_1n_1})\right\}
		\label{energy wave}
	\end{align}
	with
	\begin{align}
		&\mathcal{I}(n,m,a,b)
		\notag\\
		&=\int_0^\infty r\dd r \frac{n^2|ar^n|^2+m^2|br^m|^2+(n-m)^2|abr^{n+m}|^2}{(1+|ar^n|^2+|br^m|^2)^2} 
	\label{I}
	\end{align}
	where $n\neq m$ and $a,b\neq 0$.
	By counting the powers of $r$, one finds the integral \eqref{I} is converge if 
	\begin{alignat}{2}
	 &n>1+m>1~, &\qquad n>1,m<0~, &\qquad -1>n>m~,  \notag\\
	 &m>1+n>1~, &\qquad m>1,n<0~, &\qquad -1>m>n~.	\notag
	\end{alignat}
	We represent combinations of winding numbers for (in)finite energy solutions \eqref{energy wave} in Fig.5.

	The energy \eqref{energy wave} is associated with Noether charges. The model \eqref{F2 nls} possesses two $U(1)$ symmetries corresponding to the transformations
	\begin{align}
		&(u_1,u_2,u_3)\to (e^{i\alpha_1}u_1, ~ u_2, ~ e^{-i\alpha_1}u_3) ,
		\\
		&(u_1,u_2,u_3)\to (u_1,~ e^{i\alpha_2}u_2, ~ e^{i\alpha_2}u_3)
	\end{align} 
	where $\alpha_i, i=1,2$ are constant parameters. 
	The corresponding Noether currents are given by
	\begin{align}
		&\begin{split}
		\mathcal{J}_{(1)}^\mu=&
		\frac{4i}{\Delta_1^2}
		\left[
		(1+|u_2|^2)u_1\partial^\mu u_1^*
		-|u_1|^2u_2\partial^\mu u_2^*\right]
		\\
		&
		-\frac{4i}{\Delta_2^2}
		\left[
		(1+|u_1u_3-u_2|^2)u_3\partial^\mu u_3^*
		\right.
		\\
		&\qquad\qquad\left.
		-|u_3|^2(u_1u_3-u_2)\partial^\mu (u_1^*u_3^*-u_2^*)\right]
		\\
		&
		-\frac{i}{\Delta_1\Delta_2}
		u_1u_3(u_3^*\partial^\mu u_1^*-\partial u_2^*)
		\\
		&-\left\{u_i\leftrightarrow u_i^*\right\},
		\end{split}
		\\
		&\begin{split}
		\mathcal{J}_{(2)}^\mu=&
		\frac{4i}{\Delta_1^2}
		\left[
		(1+|u_1|^2)u_2\partial^\mu u_2^*
		-|u_2|^2u_1\partial^\mu u_1^*\right]
		\\
		&
		+\frac{4i}{\Delta_2^2}
		\left[
		(1+|u_3|^2)(u_1u_3-u_2)\partial^\mu (u_1^*u_3^*-u_2^*)
		\right.
		\\
		&\qquad\qquad\left.
		-|u_1u_3-u_2|^2u_3\partial^\mu u_3^*\right]
		\\
		&
		+\frac{i}{\Delta_1\Delta_2}
		u_2(u_3^*\partial^\mu u_1^*-\partial u_2^*)
		\\
		&-\left\{u_i\leftrightarrow u_i^*\right\}.
		\end{split}
	\end{align}
	The Noether charges per unit length are obtained as
	\begin{equation}
		Q_{(i)}=\int \dd x^2\mathcal{J}_{(i)}^0\,.
	\end{equation}
	Then for the time-dependent solutions \eqref{wave sol}, 
	the energy \eqref{energy wave} coincides with 
	\begin{equation}
		E=2\pi Q+k\left(n_1Q_{(1)}+n_2Q_{(2)}\right).
	\end{equation}
	
	It is worth to comment on scaling property like Derrick's theorem for the traveling wave solutions \eqref{wave sol}.	
	Because the traveling wave solutions localize only on the $x_1x_2$ plane, 
	we consider the rescaling $x_i\to \lambda x_i$, $i=1,2$.
	Then, the energy contribution from the $x_3$-derivative term is scaled like a potential term.
	On the other hand, as discussed in the $Q$-lump case \cite{Ward:2003un}, the contribution from the time-derivative term behaves
	like a quadratic term in $x_i$-derivatives. 
	The scaling property of the terms are opposite, 
	but when $\lambda=1$, the contribution from the $x_3$-derivative term and the time-derivative term are equal. 
	Therefore, the traveling wave solutions satisfy the Derrick type condition.
	
	
	\begin{center}
		\begin{figure}[t!]
			\includegraphics[width=7.8cm]{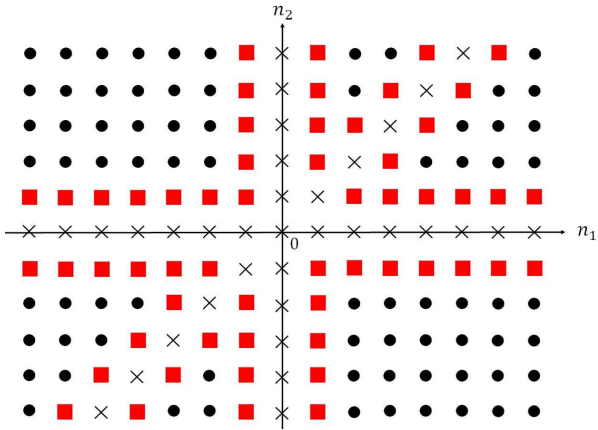}
			\caption{Combination of winding numbers corresponding to the solutions in the form \eqref{wave sol},  
			where the infinite energy (per unit length) solutions are represented by red squares and 
			black dots correspond to finite energy solutions. 
			Points with no genuine solutions are shown by cross marks.  	}
			\label{}
		\end{figure}
	\end{center}
	

	\section{Summary and Discussion}
	
	In this paper, we have constructed both static and time-dependent solutions of codimension two in the $F_2$ NL$\sigma$ model by analytically solving the second order Euler-Lagrange equation.
	The static solutions possess novel features; 
	the solutions saturate an energy lower bound, and satisfy coupled first order differential equations corresponding the saturation condition, 
	then in this sense they are BPS solutions.
	On the other hand, caused by the fact that a non-topological term is contained in the energy bound, they are saddle point solutions, i.e. sphalerons.
	As far as we know, our solutions are a rare example of "BPS sphalerons".   
	The time-dependent solutions are given by a product of localizing component on the $x^1x^2$ plane and traveling wave being parallel to $x^3$ with speed of light.
	The energy of the wave solutions are given by the topological charge and the Noether charges associated to two $U(1)$ transformations.
	Both the static and wave solutions consist of a triple in the tower of the $CP^2$ solutions generated through the \Backlund transformations, i.e. a holomorphic, anti-holomorphic and harmonic map, and therefore can be recognized as a complex of the $CP^2$ instantons and anti-instantons.
	We are currently investigating the generalization to the $F_N$ case.
	Moreover, the static solution may be able to generalize the $F_2$ model on the cylinder $\mathbb{R}\times S^1$ with twisted boundary conditions by means of a conformal mapping technique \cite{Bolognesi:2013tya}. 
	
	Since we have found several analytical solutions of the model, it is certainly worth to discuss such existence
	in the context of integrable theories in higher dimensions based on discussion of the integrable submodels 
	possessing an infinite number of local conservation laws~
	\cite{Alvarez:1997ma,Ferreira:1998zx}
	as a generalization of the previous work \cite{Bykov:2014efa,Bykov:2016rdv},
	which study integrability of two dimensional NL$\sigma$ models on homogeneous spaces.
	
	Existence and stability of the solutions which we obtained implies that instanton-anti instanton creation/annihilation may occur, through the quantum tunneling effect, in condensed matters described by
	the $SU(3)$ antiferromagnetic Heisenberg model such as cold atoms in an optical lattice \cite{Honerkamp}, and rotating dense quark matter \cite{Kobayashi:2013axa}.
	In addition, we hope this work gives a hint for finding finite energy nonembedding solutions of the $SU(3)$ pure Yang-Mills theory on the four dimensional Euclidean space.

	Apart from the BPS and also the analytical study, 
	it is also important to discuss solutions of a model which breaks the scale 
	invariance by a higher derivative order term 
	and a potential term. Such {\it baby-skyrmion} solutions may be more easy to find
	in several physical applications. 
	We will report results of these issues in forthcoming article.

	\vskip 0.5cm\noindent
	{\bf Acknowledgment}
	
	Thank you for anonymous referee for his/her careful	reading and useful comments. 
	The authors would like to Prof. L. A. Ferreira for many helpful discussions 
	and the kind hospitality at his institute where part of this work was done.
	We are very grateful to D. Bykov, S. Krusch and R. Sasaki for careful reading of the manuscript and giving helpful comments.
	We are also very grateful M. Nitta, T. Misumi, K. Toda,  H. T. Ueda and W. Zakrzewski for valuable comments and discussions.


		

\end{document}